\documentclass[%
 twocolumn,
 amsfonts,amsmath,amssymb,
 aps,
prd,
floatfix,
showpacs,
longbibliography
]{revtex4-2}

\usepackage{graphicx}% Include figure files
\usepackage{dcolumn}% Align table columns on decimal point
\usepackage{bm}% bold math
\usepackage{wasysym}% Includes astro symbols
\usepackage[dvipsnames]{xcolor}
\definecolor{ReflexBlue}{rgb}{ .0902,.0902,.5882}
\usepackage{hyperref}% add hypertext capabilities
\hypersetup{
    colorlinks=true,
    linkcolor=ReflexBlue,
    filecolor=magenta,      
    urlcolor=ReflexBlue,
    citecolor=ReflexBlue,
}
\let\Phi\varPhi

\newcommand\ba{\begin{eqnarray}}

\newcommand\ea{\end{eqnarray}}
\newcommand{\tol}{{\mbox{\tiny Tol}}}
\newcommand{\imp}{{\mbox{\tiny mod}}}
\newcommand{\cri}{{\mbox{\tiny cr}}}
\newcommand{\epow}[1]{e^{#1}}
\newcommand{\dd}{d}

\begin{document}

\title{Dynamical stability of the modified Tolman VII solution}

\author{Camilo Posada}%\email{camilo.posada@physics.slu.cz}
\affiliation{Research Centre for Theoretical Physics and Astrophysics, Institute of Physics, Silesian University in Opava, Bezru\v{c}ovo n\'{a}m. 13, CZ-746 01 Opava, Czech Republic}

\author{Jan Hlad\'ik}%\email{jan.hladik@physics.slu.cz}
\affiliation{Research Centre for Theoretical Physics and Astrophysics, Institute of Physics, Silesian University in Opava, Bezru\v{c}ovo n\'{a}m. 13, CZ-746 01 Opava, Czech Republic}

\author{Zden\v{e}k Stuchl\'ik}%\email{zdenek.stuchlik@physics.slu.cz}
\affiliation{Research Centre for Theoretical Physics and Astrophysics, Institute of Physics, Silesian University in Opava, Bezru\v{c}ovo n\'{a}m. 13, CZ-746 01 Opava, Czech Republic}
\date{\today}

\begin{abstract}
Studies of neutron stars are at their peak after the multi-messenger observation of the binary merger event GW170817, which strongly constraints the stellar parameters like tidal deformability, masses and radii. Although current and future observations will provide stronger limits on the neutron stars parameters, knowledge of explicit interior solutions to Einstein's equations, which connect observed parameters with the internal structure, are crucial to have a satisfactory description of the interior of these compact objects. A well known exact solution, which has shown a relatively good approximation to a neutron star, is the Tolman VII solution. In order to provide a better fitting for the energy density profile, with the realistic equations of state for neutron stars, recently Jiang and Yagi proposed a modified version of this model which introduces an additional parameter $\alpha$ reflecting the interplay of the quadratic and the newly added quartic term in the energy density profile. Here we study the dynamical stability of this modified Tolman VII solution using the theory of infinitesimal and adiabatic radial oscillations developed by Chandrasekhar. For this purpose, we determine values of the critical adiabatic index, for the onset of instability, considering configurations with varying compactness and $\alpha$. We found that the new models are stable against radial oscillations for a considerable range of values of compactness and the new parameter $\alpha$, thus supporting their applicability as a physically plausible approximation of realistic neutron stars.
\end{abstract}

\maketitle

\section{Introduction}
\label{intro}
Current studies of neutron stars (NSs) can offer important insight on the properties of cold, catalized matter with densities beyond the nuclear saturation regime, $n_{s}\sim 0.16\,\mathrm{fm}^{-3}$, which cannot be investigated via heavy ion collisions \cite{Lattimer:2019eez}. Astrophysical observations of NS can provide, in principle, valuable constraints on the equation of state (EOS), i.e, the relation between pressure and density, which remains one of the biggest unknowns in relativistic astrophysics. The most well-studied parameters of NS are the tidal deformability \cite{Abbott:2017qsa,Abbott:2018wiz}; and the mass and radius \cite{Lattimer:2019eez}. The current mission, \emph{Neutron Star Interior Composition Explorer} NICER \cite{Arzoumanian:2014} is providing estimates of mass and radius of NSs, thus putting tight constraints on the realistic EOSs \cite{Riley:2019yda}.

Besides the constraints on astrophysical parameters of NSs, it is crucial to obtain interior solutions to Einstein's equations which connects observables (tidal deformability, mass, radius) with the internal structure of the configuration. Finding analytical stellar interior solutions can be an arduous task given the complexity of the field equations. However, some few exact solutions have been used in the literature to model NS. The simplest is the constant-density Schwarzschild interior solution\footnote{There are various reasons for considering uniform density configurations. See for instance \cite{Harrison:1965}, Ch. 8} \cite{Schwarzschild:1916,Stuchlik:2000}, which corresponds to the particular case of a polytropic sphere with $n=0$ \cite{Tooper:1964,Stuchlik:2016xiq}. Another solution is the one found by Buchdahl \cite{Buchdahl:1967ApJ} which turned out to be stable against radial oscillations \cite{Negi:2007GRG,Moustakidis:2016ndw}. The most popular one, however, is the so-called Tolman VII solution \cite{Tolman:1939jz}. This is a two-parameter solution (mass and radius), which is characterised by a vanishing density and pressure at the surface. This solution was found to be dynamically stable \cite{Moustakidis:2016ndw,Negi:1999grg,Negi:2001ApS,Raghoonundun:2015wga}, and it also exhibits trapped null geodesics for $R/M>3$ \cite{Neary:2001ai}.

Recently a modified version of the Tolman VII solution was proposed by \cite{Jiang:2019vmf} where a new parameter $\alpha$ was introduced, such that the energy density becomes a combination of quadratic and quartic functions of the radial coordinate $r$. It was shown that the new parameter $\alpha$ can be expressed in terms of the stellar parameters -- mass $M$, radius $R$ and central energy density $\epsilon_\mathrm{c}$ -- insensitive of the EOS. Thus, this modified Tolman VII solution is characterised by three parameters ($M$, $R$, $\epsilon_\mathrm{c}$). Due to the complexity of the modified energy density profile, the authors were not able to find an exact analytic solution to Einstein's equations, therefore this new model remains as an approximation. Nevertheless, this modified solution seems to model more accurately realistic profiles of NSs than the original Tolman VII. The I-Love-C relations for the modified Tolman VII solution were studied in \cite{Jiang:2020uvb}. 

In this paper we study the dynamical stability, under radial perturbations, of spherically symmetric fluid spheres described by the modified Tolman VII model. We use the variational method developed by Chandrasekhar \cite{Chandrasekhar:1964zz,Bardeen:1966} to investigate the general constraints on the stellar parameters in order to construct solutions dynamically stable. For this purpose, we determine the critical adiabatic index, for the onset of instability, for several values of the compactness $M/R$ and the parameter $\alpha$. Our results provide relevant constraints for the physical plausibility of the modified Tolman VII solution. 

The paper is organised as follows. In Sec.~\ref{sec:2} we summarize the methodology used by Tolman to obtain interior solutions for relativistic fluid spheres; we review the original Tolman VII solution and the modified version proposed by \cite{Jiang:2019vmf}. In Sec.~\ref{sec:3} we present the Chandrasekhar variational method to study the radial stability of relativistic spherical masses. In Sec.~\ref{sec:4} we present our results. Final conclusions are discussed in Sec.~\ref{sec:5}. Throughout the paper we use geometric units, $c=G=1$.

%************************************************************************************************
%************************************************************************************************

\section{Tolman's method for the solution of a fluid in equilibrium}
\label{sec:2}

Following Tolman \cite{Tolman:1939jz}, we consider a static and spherically symmetric matter distribution. Thus, we choose the line element ansatz in the standard Schwarzschild-like form
\ba\label{metric}
ds^2=-e^{\nu}\dd t^2 + e^{\lambda}\dd r^2 + r^2\left(\dd\theta^2 + \sin^2\theta\dd\phi^2\right),
\ea 
\noindent where $\nu$ and $\lambda$ depend only on $r$. We assume that the matter inside the configuration is described by a perfect fluid; its stress-energy tensor satisfies the general form
 \ba\label{tmunu}
T_{\mu\nu}=(\epsilon+p)u_{\mu}u_{\nu}+pg_{\mu\nu}.
\ea 
Here $\epsilon$ indicates the energy density, $p$ is the pressure, and $u^{\mu}=\dd x^{\mu}/\dd s$ is the four-velocity. 

Substituting Eqs.~\eqref{metric} and \eqref{tmunu} into Einstein's equations $G_{\mu\nu}=8\pi T_{\mu\nu}$, one finds \cite{Tolman:1939jz}
\ba\label{einstein1}
\frac{\dd}{\dd r}\left(\frac{e^{-\lambda}-1}{r^2} + \frac{e^{-\lambda}\nu'}{2r}\right)+e^{-(\lambda+\nu)}\frac{\dd}{\dd r}\left(\frac{e^{\nu}\nu'}{2r}\right)=0,
\ea 
\ba\label{einstein2}
e^{-\lambda}\left(\frac{\nu'}{r}+\frac{1}{r^2}\right) - \frac{1}{r^2}=8\pi p,
\ea
\ba\label{einstein3}
e^{-\lambda}\left(\frac{\lambda'}{r}-\frac{1}{r^2}\right) + \frac{1}{r^2}=8\pi\epsilon.
\ea
We have three differential equations for the unknown functions $\nu$, $\lambda$, $p$ and $\epsilon$. Once an equation of state (EOS) $p(\epsilon)$, connecting pressure with energy density, is provided, the system will be determinate. It is conventional to define the mass $m(r)$ enclosed in the radius $r$ as
\ba\label{einstein4}
e^{-\lambda(r)}\equiv1-\frac{2m(r)}{r}\,.
\ea
\noindent In a more mathematical rather than ‘physical' approach, which turned out to be more propitious to integrate the system of equations \eqref{einstein1}-\eqref{einstein3}, Tolman \cite{Tolman:1939jz} chose conveniently certain relations for $\nu$ and $\lambda$, or both, as a function of $r$, and then he analysed the physical plausibility of the solutions obtained. Following this approach, Tolman re-derived the Schwarzschild interior solution, the Einstein universe, among others. 

In the next subsection we will discuss one of the solutions obtained by Tolman, using the method described above, known as the Tolman VII solution. 

%************************************************************************************************
%************************************************************************************************

\subsection{Tolman VII solution}\label{sec:1a}

\label{sec:1a}
In the following we adopt the convention used in \cite{Jiang:2019vmf}. In this section we briefly summarize the analytic solution to Einstein's equations discovered by Tolman \cite{Tolman:1939jz}, which we will refer to as T-VII. Tolman assumed $e^{-\lambda(r)}$ in the form \footnote{Henceforth the subscript ‘Tol' will indicate quantities associated to the original T-VII solution}
\ba\label{grrTol}
e^{-\lambda(x)_{\tol}}=1-\mathcal{C}x^2 (5-3x^2),
\ea 
\noindent where $x\equiv r/R$, with $R$ denoting the radius of the configuration, and $\mathcal{C}\equiv M/R$ is the compactness. Under this assumption, the energy density $\epsilon(r)$, mass $m(r)$ and pressure $p(r)$ are given by
\ba\label{rhoTol}
\epsilon(x)_{\tol} = \epsilon_\mathrm{c}(1-x^2),\quad m_\tol(x) &=& \frac{M}{2}x^3 \left(5 - 3x^2\right);
\ea
\ba\label{pTol}
\frac{p_{\tol}}{\epsilon_{\mathrm{c}}}=\frac{1}{15}\left[\sqrt{\frac{12e^{-\lambda_{\tol}}}{\mathcal{C}}}\tan\phi_{\tol}-(5-3x^2)\right].
\ea
Here $\epsilon_{\mathrm{c}}$ is the central energy density and $M=m(R)$ is the total stellar mass. In terms of the radius $R$ and central energy density $\epsilon_{\mathrm{c}}$, the  compactness can be written as
\ba\label{CTol}
\mathcal{C}=\frac{8\pi}{15}\epsilon_{\mathrm{c}}R^2.
\ea 
Note that the energy density vanishes at the boundary $r=R$. The $g_{tt}$ metric component results
\ba\label{gttTol}
e^{{\nu(r)}_{\tol}}=C_{1}^{\tol}\cos^2 \phi_{\tol}\,,
\ea
\noindent where
\ba\label{phiTol}
\quad \phi_{\tol}=C_{2}^{\tol}-\frac{1}{2}\log\left(x^2-\frac{5}{6}+\sqrt{\frac{5e^{-\lambda_{\tol}}}{8\pi\epsilon_{\mathrm{c}}R^2}} \right).
\ea
\noindent Here $C_{1}^{\tol}$ and $C_{2}^{\tol}$ are constants of integration \cite{Tolman:1939jz}. It is worthwhile to recall certain restrictions for the physical plausibility of the T-VII solution. For instance, from Eq.~\eqref{pTol} we find that the central pressure diverges when $\mathcal{C}=0.3862$. Note that this limit satisfies the general inequality $\mathcal{C}_{B}\leq 4/9$ derived by Buchdahl \cite{Buchdahl:1959} for general relativistic static fluid spheres.

An additional restriction, which was not considered in the literature \cite{Moustakidis:2016ndw,Negi:1999grg,Raghoonundun:2015wga,Neary:2001ai}, is determined by the dominant energy condition (DEC) which holds for all kinds of matter, including electromagnetic and scalar fields. The DEC states that for a perfect fluid the energy density must be nonnegative and greater or equal to the magnitude of the pressure $\epsilon\geq\vert p\vert$ \cite{Carroll:2019st}. We found that for the T-VII solution, $p_{c}<\epsilon_{\mathrm{c}}$ for $\mathcal{C}<0.3351$. Note that this limit is lower than the one provided by the condition of finite central pressure discussed above. Therefore the T-VII solution violates the DEC in the range $\mathcal{C}\in(0.3351,0.3862)$.

\begin{figure}[htb]
   \includegraphics[width=\columnwidth]{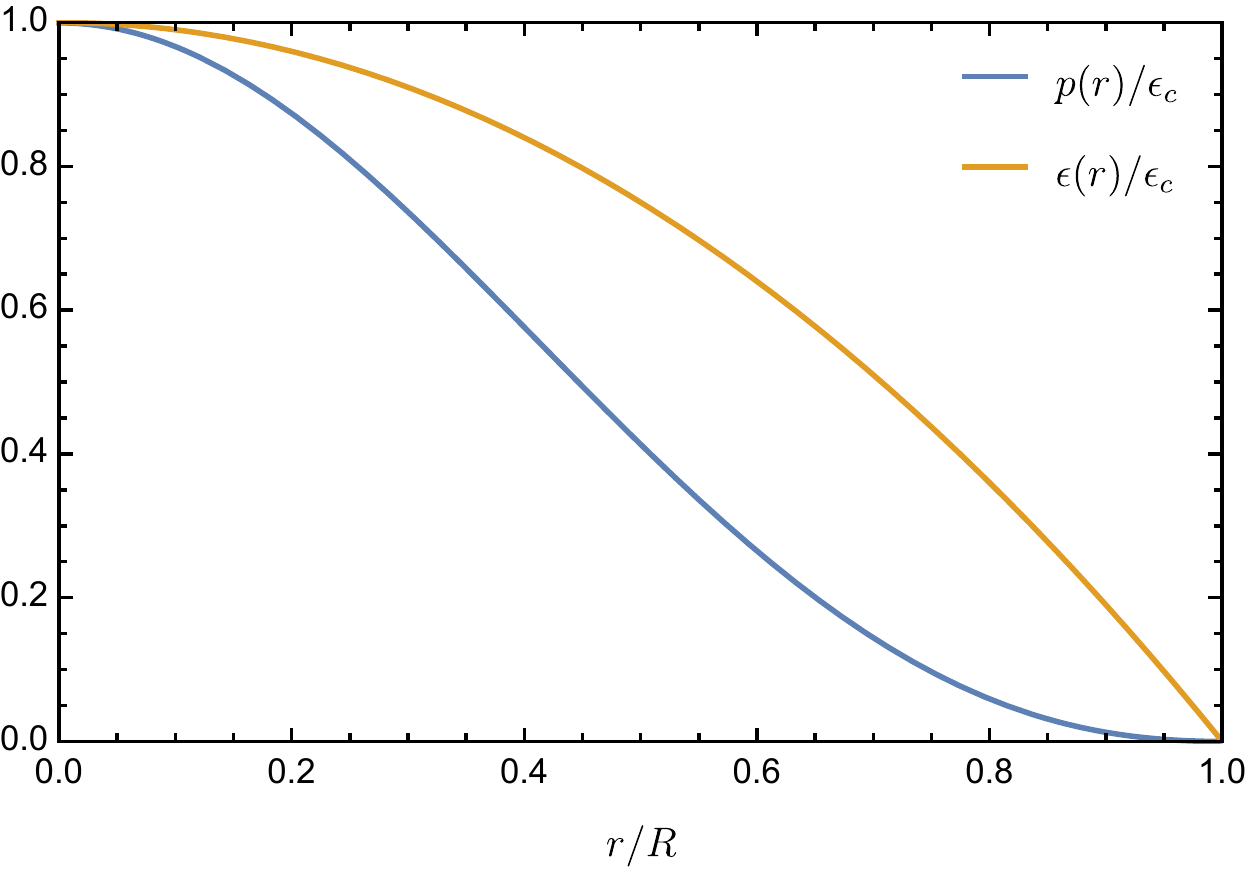}
   \caption{\label{fig1} Profiles of the energy density and pressure, in units of the central energy density $\epsilon_\mathrm{c}$, as a function of the dimensionless ratio $r/R$, for the T-VII solution. We show profiles for a configuration with $\mathcal{C}=0.3351$, corresponding to the limit given by the DEC.}
    \label{fig:1}
\end{figure}

%************************************************************************************************
%************************************************************************************************

\subsection{Modified Tolman VII solution}\label{sec:1b}

 In this subsection we review the modified T-VII solution proposed in \cite{Jiang:2019vmf}. In this new model, the energy density $\epsilon$ is assumed in the polynomial form
\ba\label{rhoMod}
\epsilon_{\imp}=\epsilon_{\mathrm{c}}\left[1-\alpha x^2 + (\alpha-1)x^4\right],
\ea
\noindent where $\alpha$ is a new free parameter of the solution. In principle, this modified T-VII solution seems to model more accurately the energy density profile for realistic EOS of NSs, as compared with the original T-VII solution. When $\alpha=1$, Eq.~\eqref{rhoMod} reduces to the original T-VII energy density [Eq. \eqref{rhoTol}]. The remaining expressions of the modified T-VII solution are the following \cite{Jiang:2019vmf}
\ba\label{grrMod}
e^{-\lambda_{\imp}}=1-8\pi\epsilon_{\mathrm{c}} R^2 x^2 \left[\frac{1}{3}-\frac{\alpha}{5}x^2 + \frac{(\alpha-1)}{7}x^4 \right],
\ea 
\ba
e^{{\nu}_{\imp}}=C_1^\imp \cos^2 \phi_{\imp},
\ea
\ba\label{mMod}
m_\imp (x) =4 \pi \epsilon_\mathrm{c} R^3 x^3 \left(\frac{1}{3} - \frac{\alpha}{5}x^2 +\frac{\alpha - 1}{7} x^4\right),
\ea
\ba\label{pMod}
\frac{p_{\imp}(x)}{\epsilon_{\mathrm{c}}} &=& \left(\frac{e^{-\lambda_{\tol}}}{10\pi\epsilon_\mathrm{c} R^2}\right)^{1/2}\tan\phi_{\imp}+\frac{1}{15}(3x^2-5)\nonumber\\
&&+\frac{6(1-\alpha)}{16\pi\epsilon_\mathrm{c} R^2(10-3\alpha) -105},
\ea
\noindent where,
\ba
\phi_\imp &=& C_2^\imp - \frac{1}{2} \log\left(x^2 - \frac{5}{6} +\sqrt{\frac{5 e^{-\lambda_\tol}}{8 \pi R^2 \epsilon_\mathrm{c}}}\right).
\ea
Here $C_1^\imp$ and $C_2^\imp$ are integration constants \cite{Jiang:2019vmf}. Note that some of Einstein's equations are fulfilled exactly, while the others are satisfied only approximately \cite{Jiang:2019vmf}. The modified T-VII is a four-parameter model, namely, $(\epsilon_{\mathrm{c}},\mathcal{C},R,\alpha)$. Using Eq.~\eqref{mMod} and considering that $M=m_{\imp}(1)$, we can eliminate the radius $R$ 
\ba
8\pi\epsilon_\mathrm{c} R^2 = \frac{105\mathcal{C}}{10-3\alpha}.
\ea
Thus, the modified T-VII model reduces to a three-parameter $(\epsilon_{\mathrm{c}},\mathcal{C},\alpha)$ solution. We will follow this convention in the subsequent calculations. 
 
 \section{Dynamical stability of the modified T-VII model}
 \label{sec:3}
 
 The stability under infinitesimal and adiabatic radial perturbations, for any equilibrium configuration, can be rigorously studied via the eigenequation derived by Chandrasekhar \cite{Chandrasekhar:1964zz}. We follow the prescription of \cite{Misner:1974qy}, where the radial motion of the configuration is written in terms of the variable $\xi$, which represents the ‘Lagrangian displacement' from equilibrium
\ba\label{amplitude}
\xi=r^{-2} e^{\nu/2}\zeta,
\ea
where $\zeta$ is the so-called ‘renormalized displacement function' assumed to be in the form $\zeta=\zeta(r)e^{-i\omega t}$, with $\omega$ denoting the frequency of the oscillations. The ‘pulsation' equation reduces to the Sturm-Liouville form for $\zeta(r)$ as follows \cite{Misner:1974qy}
\ba\label{pulsation}
\frac{\dd}{\dd r}\left[P\left(\frac{\dd\zeta}{\dd r}\right)\right] + (Q+\omega^{2} W)\zeta=0.
\ea
The functions $P(r)$, $Q(r)$ and $W(r)$ are given in terms of the variables of the fluid in equilibrium
\ba\label{Pr}
P(r)\equiv  \frac{\gamma\, p}{r^2}\,\epow{(3\nu+\lambda)/2}\, ;\quad  W(r)\equiv\frac{\epsilon + p}{r^2}\,\epow{(3\lambda+\nu)/2}\,,
\ea
\ba\label{Qr}
    Q(r) &\equiv& -4e^{(3\nu+\lambda)/2}\,
    r^{-3}\,\frac{\dd p}{dr} - 8\pi e^{3(\nu+\lambda)/2}\,r^{-2}\,p\,(\epsilon + p)\nonumber \\
    &&+e^{(3\nu+\lambda)/2}\,r^{-2}\,(\epsilon + p)^{-1}\left(\frac{\dd p}{dr}\right)^2.
\ea
The adiabatic index $\gamma$, governing the \emph{perturbations}, is given by
\ba
\gamma=\frac{\epsilon+p}{p}\left(\frac{\partial p}{\partial\epsilon}\right)_{\mathrm{ad}}=\left(1+\frac{\epsilon}{p}\right)(v_{\mathrm{s}})_{\mathrm{ad}}^2\,,
\ea
where $(v_{\mathrm{s}})_{\mathrm{ad}}=\left(\partial p/\partial\epsilon\right)^{1/2}$ is the speed of sound (in units of the speed of light). The subscript “ad" indicates that the derivative is taken for an adiabatic process. Let us remark that this adiabatic index $\gamma$ does not necessarily equals the adiabatic index $\Gamma$ associated to the equilibrium pressure-density relation (see e.g. \cite{Shapiro:1983du} for a discussion). 

Physically acceptable solutions to the eigenequation \eqref{pulsation} must satisfy the following boundary conditions. Equation \eqref{pulsation} diverges at the origin $r=0$, but $\xi$ and $\dd\xi/dr$ must be finite there, so we are led to
\ba\label{bc1}
\zeta \sim r^3,\quad \mathrm{as}\quad r\to 0.
\ea 
Condition \eqref{bc1} states that there are no displacements of the fluid at the center. On the other hand, the Lagrangian change in pressure at the surface $r=R$ must vanish, i.e., $\Delta p=0$, which implies
\ba\label{bc2}
\gamma\,p\,r^{-2}e^{\nu/2}\left(\frac{\dd\zeta}{\dd r}\right) \to 0,\quad \mathrm{as}\quad r\to R.
\ea 
The ‘pulsation' equation [Eq.~\eqref{pulsation}] corresponds to an eigenvalue problem for the frequencies $\omega$ and displacements $\zeta$. The eigenequation \eqref{pulsation} can be re-expressed in a variational form \cite{Bardeen:1966,Harrison:1965}. The eigenvalue $\omega^2$ can be determined by the extremal value of the R.H.S. of 
\ba\label{variat} 
\omega^2 = \frac{\int_{0}^{R}\left[P\zeta'^2 - Q\zeta^2\right]\dd r}{\int_{0}^{R}W\zeta^2 \dd r},
\ea
where the function $\zeta(r)$, which gives the extremal value, is the corresponding eigenfunction. A common technique to solve the eigenvalue problem given by Eq.~\eqref{variat} is to vary over all functions, subject to the boundary conditions \eqref{bc1} and \eqref{bc2}, which satisfy the orthogonality relation
\ba\label{eigenf} 
\int_{0}^{R}W(r)\,\xi_{i}\,\xi_{j}\dd r=\delta_{ij},
\ea  
where $\xi_{i}$ and $\xi_{j}$ are the eigenfunctions for different characteristic eigenvalues. Thus, a sufficient condition for the instability under radial oscillations is that the R.H.S of Eq.~\eqref{variat} vanishes for certain ‘trial function' $\xi$ which satisfies the boundary conditions \eqref{bc1} and \eqref{bc2}. The various methods to solve the eigenvalue problem [Eq.~\eqref{variat}] are discussed in \cite{Bardeen:1966}.  

The neutral, or marginally stable, mode appears when $\omega^2 = 0$. Imposing this condition in Eq.~\eqref{variat}, we find the critical adiabatic index $\gamma_{\cri}$ for the onset of dynamical instability
\ba\label{gammacr} 
\gamma_{\cri}=\frac{\int_{0}^{R}Q(r)\,(r^2\,\epow{-\nu/2}\xi)^2\,\dd r}{\int_{0}^{R}\epow{(3\nu+\lambda)/2}\,r^{-2}\,p\,\left[\frac{\dd}{\dd r}(r^2\,\epow{-\nu/2}\xi)\right]^2\dd r},
\ea  
where we have used Eq.~\eqref{amplitude}. It is conventional to define, for any equilibrium configuration, an ‘effective', or averaged, adiabatic index in the form \cite{Moustakidis:2016ndw,Merafina:1989}
\ba\label{gammaeff} 
\langle\gamma\rangle=\frac{\int_{0}^{R}\epow{(3\nu+\lambda)/2}\,\gamma\,p\,r^{-2}\,\left[\frac{\dd}{\dd r}(r^2\,\epow{-\nu/2}\xi)\right]^2\dd r}{\int_{0}^{R}\epow{(3\nu+\lambda)/2}\,p\,r^{-2}\,\left[\frac{\dd}{\dd r}(r^2\,\epow{-\nu/2}\xi)\right]^2\dd r},
\ea  
thus there will be stable equilibrium when 
\ba\label{stablec}
\langle\gamma\rangle \geq \gamma_{\cri}.
\ea
The preceding discussion is the most general one, but some few special cases have been studied in the literature \cite{Moustakidis:2016ndw,Chandrasekhar:1964zz,Camilo:2018goy,Hladik:2020xfw,Posada:2020svn}. For instance, for the case of homogeneous stars with constant density, Chandrasekhar obtained the well known result \cite{Chandrasekhar:1964zz}
\ba
\gamma_{\cri}=\frac{4}{3}+\left(\frac{19}{42}\right)2\mathcal{C}+\mathcal{O}(\mathcal{C}^2).
\ea

In the next section we present our results for the dynamical stability of the original and modified T-VII solutions.

\section{Results}
\label{sec:4}
In this section we present our results for the dynamical stability of the modified T-VII solution. In our analysis we consider values of $\alpha$ in the range, $\alpha\in[0,2]$. This regime is restricted by the solution in the following way; for $\alpha<0$ the density is a non-monotonically decreasing function of $r$, which is not consistent with the realistic EOSs for NSs. On the other hand, for $\alpha>2$ the solution shows a negative energy density. Note that our range of $\alpha$-values is more extensive than the one considered by \cite{Jiang:2019vmf,Jiang:2020uvb} who restricted $\alpha\in[0.4,1.4]$, and $\mathcal{C}\in[0.05,0.35]$, which seems to be the characteristic regime for realistic NSs after fitting a numerical energy density profile with Eq.~\eqref{rhoMod} for 11 EOSs. 

In order to determine the stability domain of the modified T-VII solution, we computed the critical values of the adiabatic index $\gamma{\cri}$ [Eq.~\eqref{gammacr}], and the values of the effective adiabatic index $\langle\gamma\rangle$ [Eq.~\eqref{gammaeff}] using the following two trial functions
\ba\label{trials}
 \xi_{1}=r\epow{\nu/2};\qquad \xi_{2}=r\epow{\nu/4}\,,
\ea
which satisfy the boundary conditions \eqref{bc1} and \eqref{bc2}. These two trial functions are the most commonly used in the literature \cite{Chandrasekhar:1964zz,Merafina:1989,Moustakidis:2016ndw}. In principle, some other trial function, different from the conventionally used ones, could provide stronger limits on the compactness for stability. However there is no way to guess which trial function would be the most convenient. A power series is frequently useful in variational calculation \cite{Bardeen:1966}.

\begin{figure*}[h]
\includegraphics[width=0.49\linewidth]{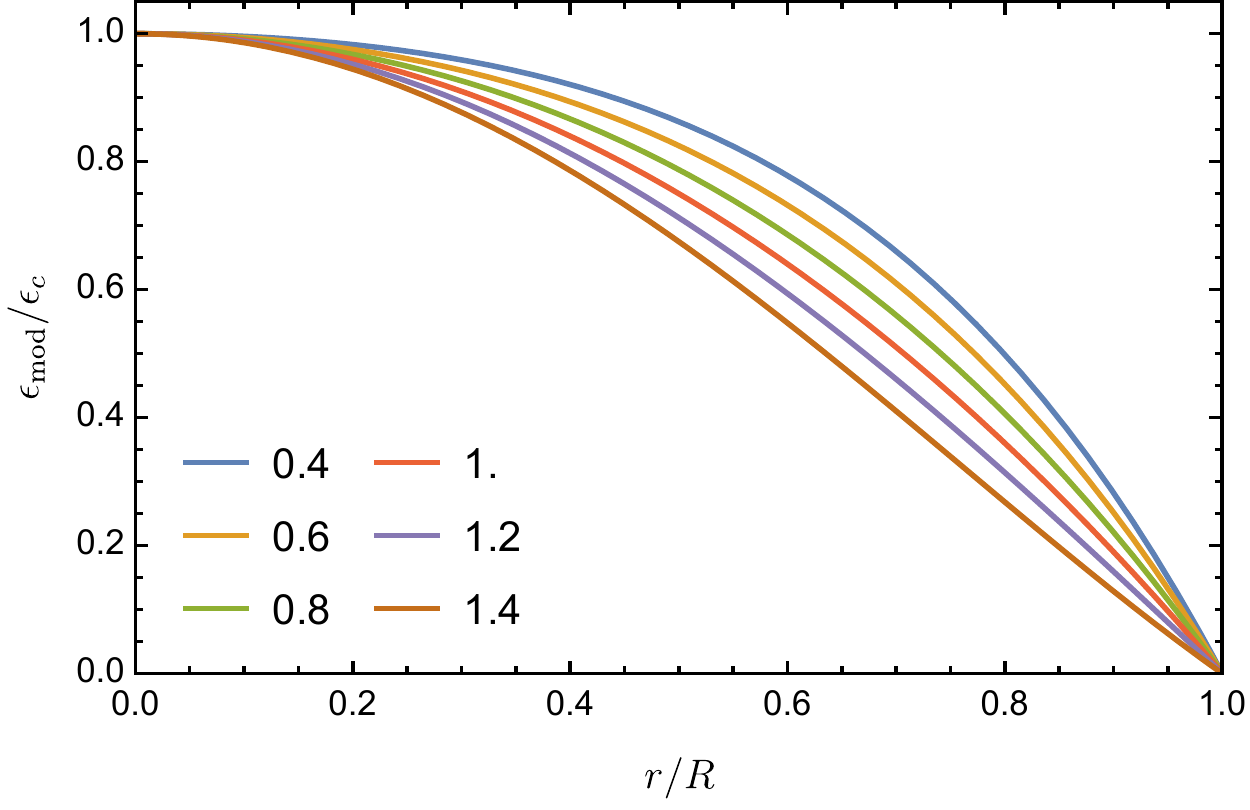}
\includegraphics[width=0.49\linewidth]{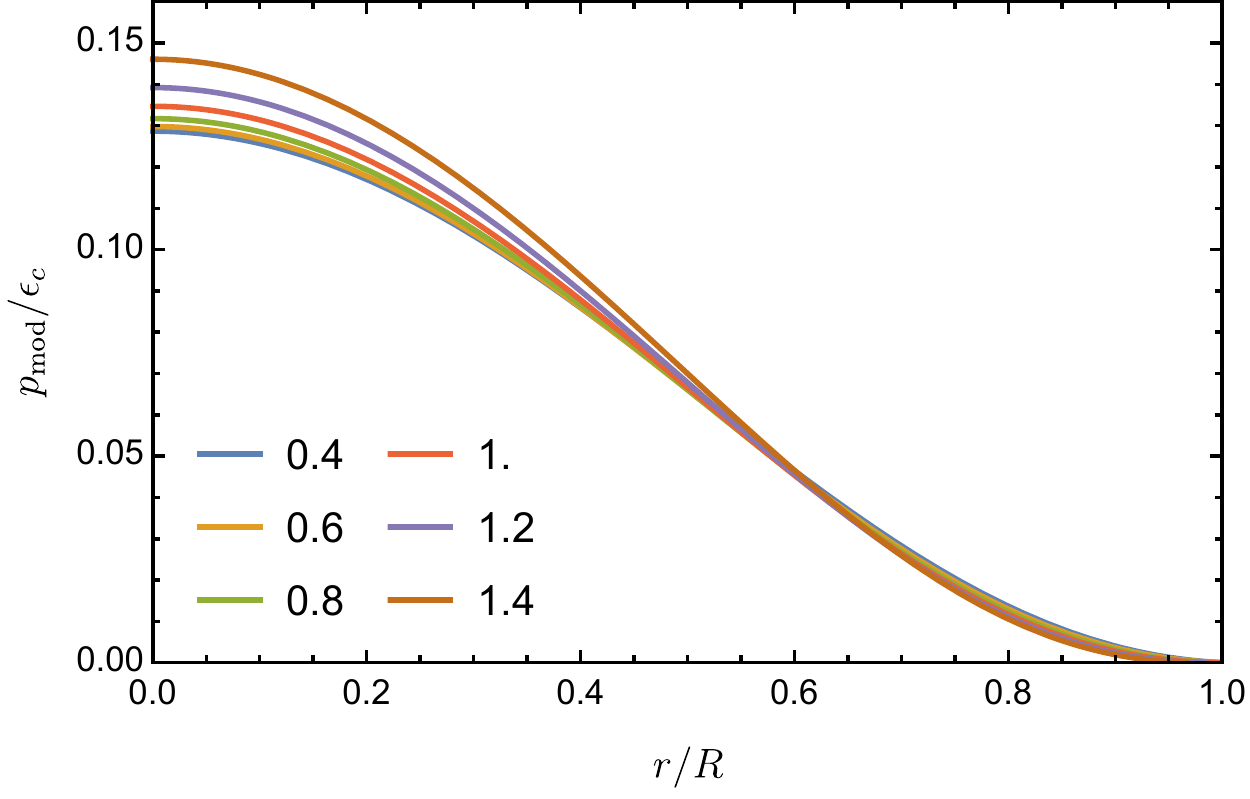}
\caption{Profiles of characteristic quantities of the modified T-VII model. {\bf Left:} Energy density (in units of $\epsilon_\mathrm{c}$) as a function of $r/R$. {\bf Right:} Pressure (in units of $\epsilon_\mathrm{c}$) as a function of $r/R$. We show profiles for some representatives values of $\alpha\in[0.4,1.4]$. The case $\alpha=1$ corresponds to the original T-VII solution. Note that the energy density vanishes at the surface.}
\label{fig:2}       % Give a unique label
\end{figure*}

Before presenting our results on the stability, let us discuss some general properties of the modified T-VII solution. In Fig.~\ref{fig:2} we present the energy density (left panel) and pressure (right panel), in units of the central energy density $\epsilon_\mathrm{c}$, for the modified T-VII model. We display profiles for several values of $\alpha$. The case $\alpha=1$ corresponds to the original T-VII solution. Note that a common feature of this model is that the energy density vanishes at the surface.
\begin{figure*}[h]
\includegraphics[width=0.49\linewidth]{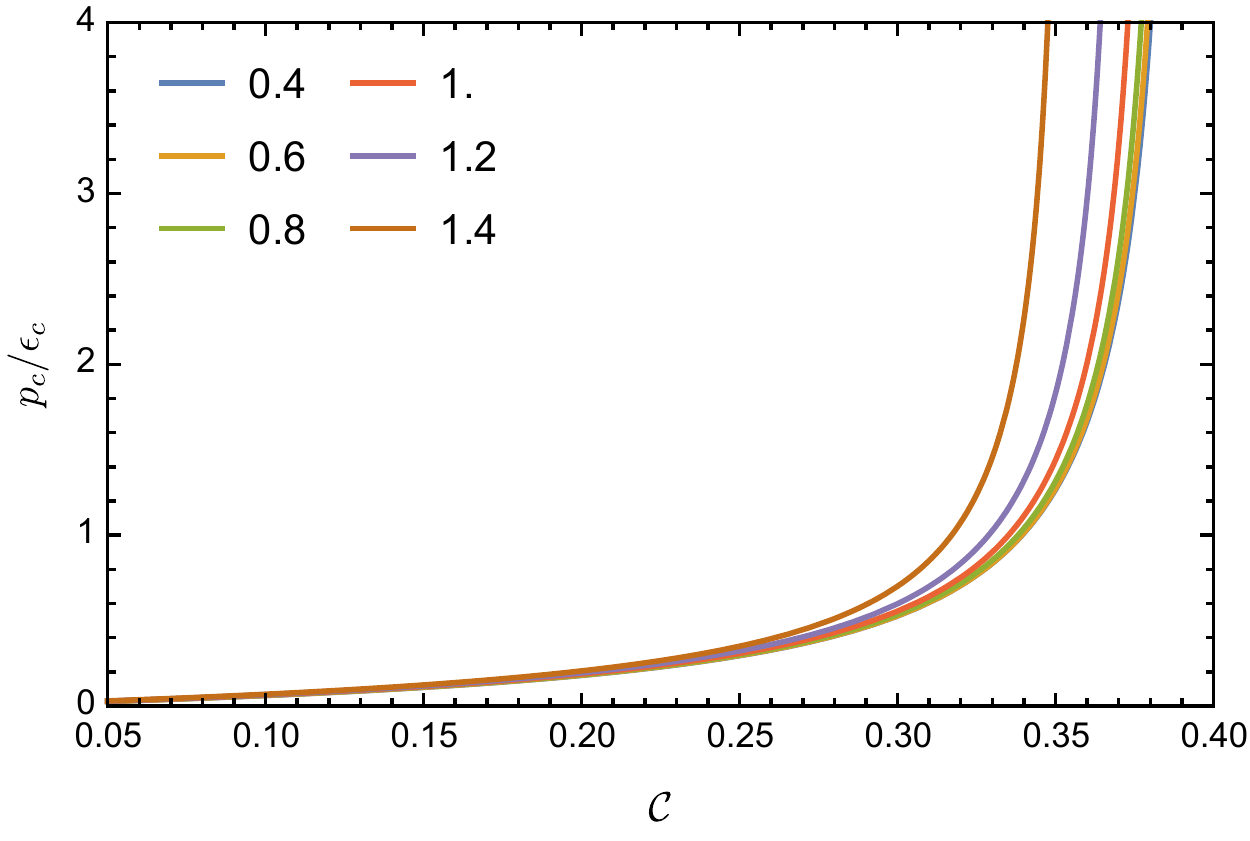}
\includegraphics[width=0.49\linewidth]{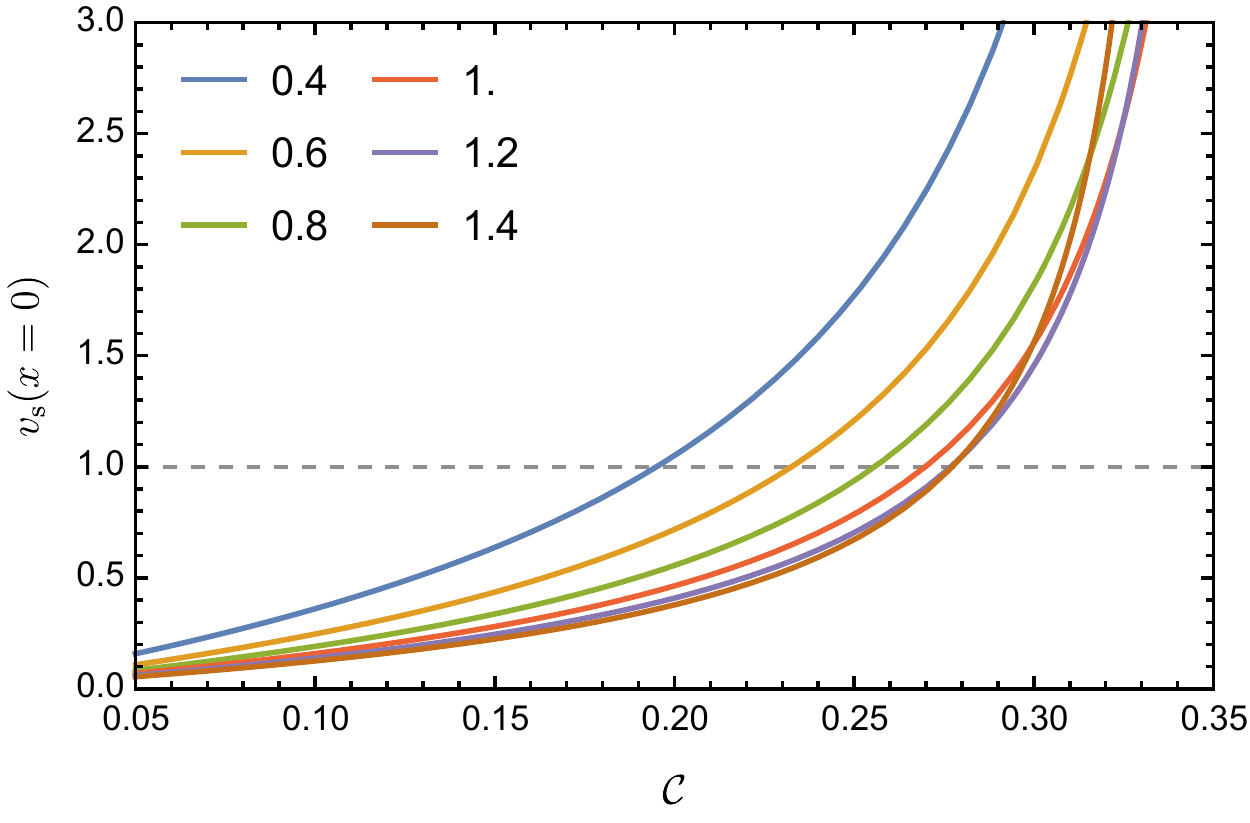}
\caption{Properties of the modified T-VII model. {\bf Left:} Central pressure (in units of $\epsilon_\mathrm{c}$) as a function of the compactness. {\bf Right:} Speed of sound at the center (in units of $c$) as a function of $\mathcal{C}$. We show profiles for the same values of $\alpha$ as in Fig.~\ref{fig:2}}
\label{fig:3}       % Give a unique label
\end{figure*}

In Fig.~\ref{fig:3} (left panel) we show the ratio of the central values of pressure and energy density $p_{c}/\epsilon_{\mathrm{c}}$, as a function of the compactness, for the modified T-VII solution. We consider the same values of $\alpha$ as in Fig.~\ref{fig:2}. Note that as $\alpha$ increases the limiting compactness for having a finite central pressure decreases.  In the same figure (right panel) we also display the speed of sound measured at the center (in units of the speed of light), as a function of $\mathcal{C}$. Note that the limits on the compactness set by causality are lower than those determined by the condition of finite central pressure.

\begin{figure*}%[htb]
\includegraphics[width=0.49\linewidth]{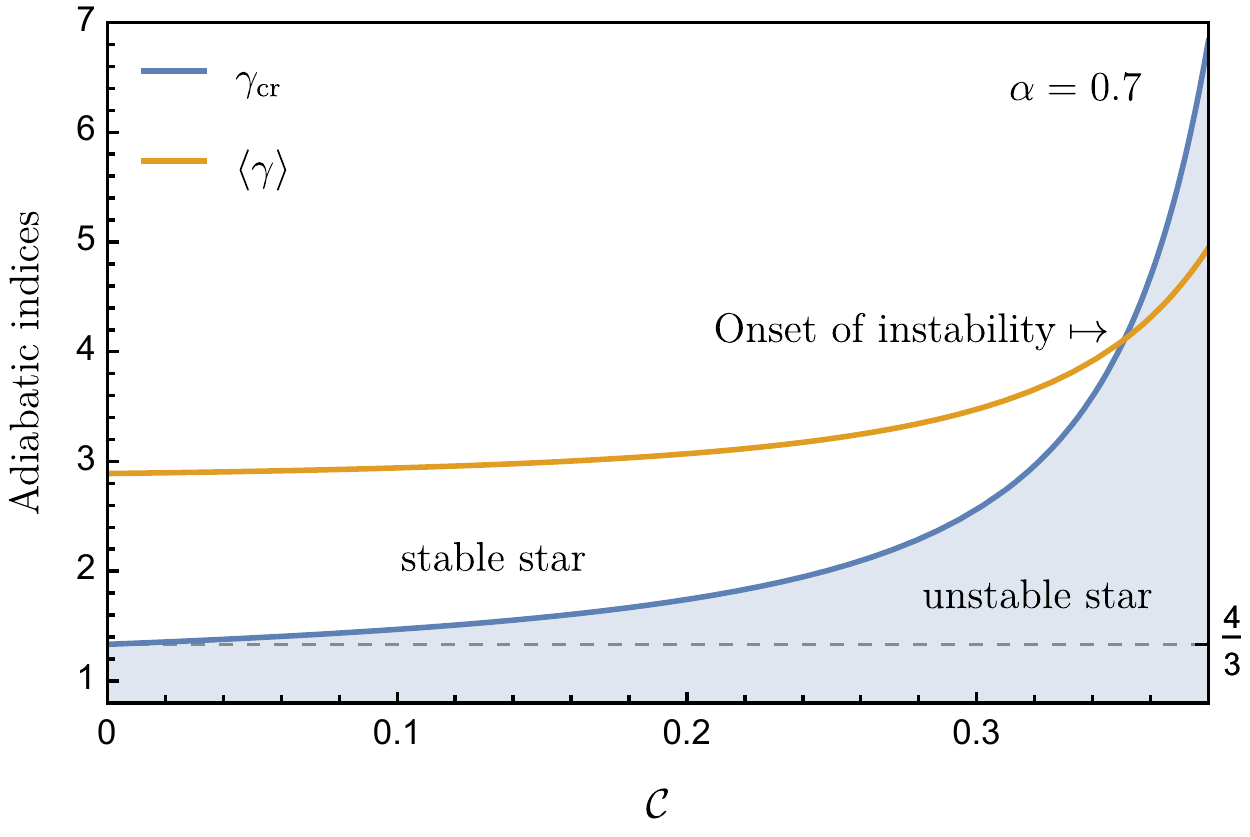}
\hfill
\includegraphics[width=0.49\linewidth]{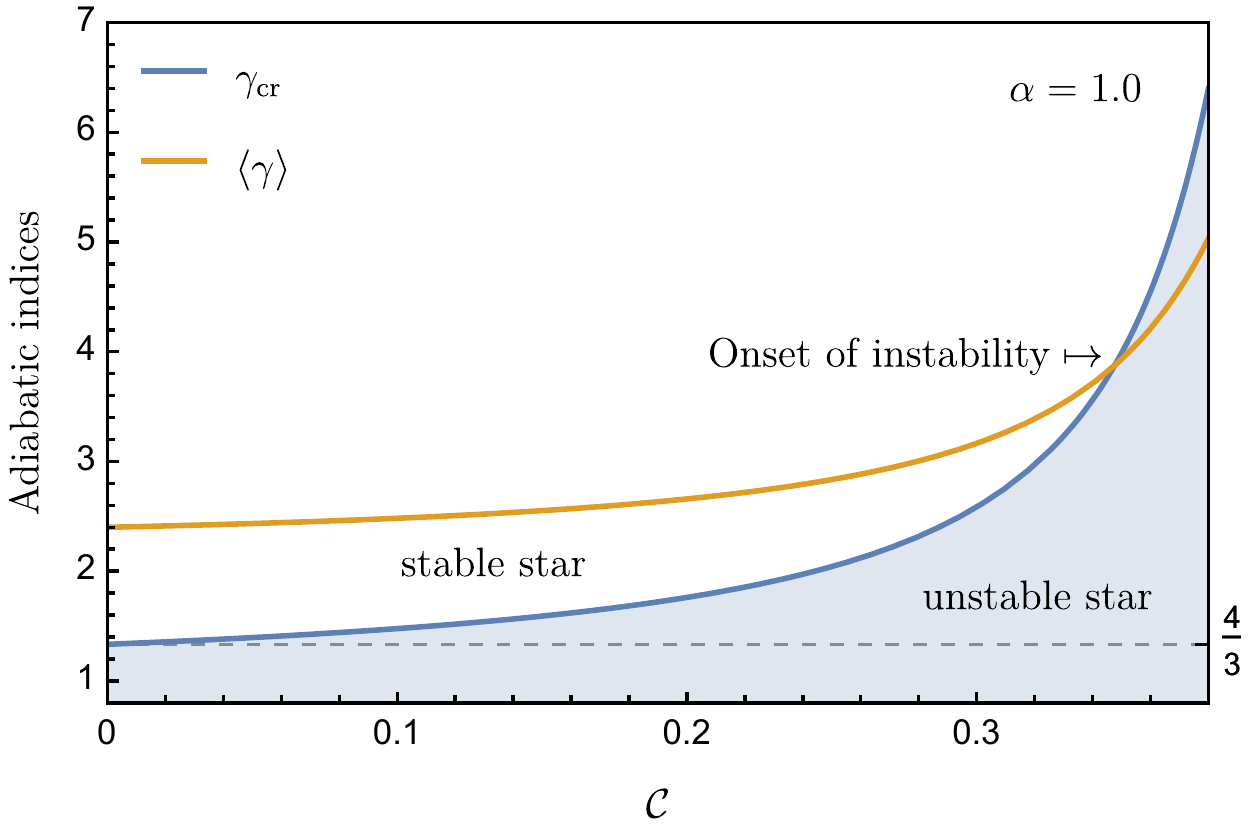}
\caption{
\label{fig:4} The stability domain for the modified T-VII model for $\alpha=0.7$ (left panel) and the original T-VII solution (right panel), as determined by comparison of the effective and critical adiabatic indices. The values of $\gamma_{\cri}$ were computed via the trial function $\xi_2$. The stability regime is determined by the condition $\langle\gamma\rangle > \gamma_{\cri}$. The intersection point marks the limiting value of the compactness for stability. }
\end{figure*}

\begin{figure}[htb]
\includegraphics[width=\columnwidth]{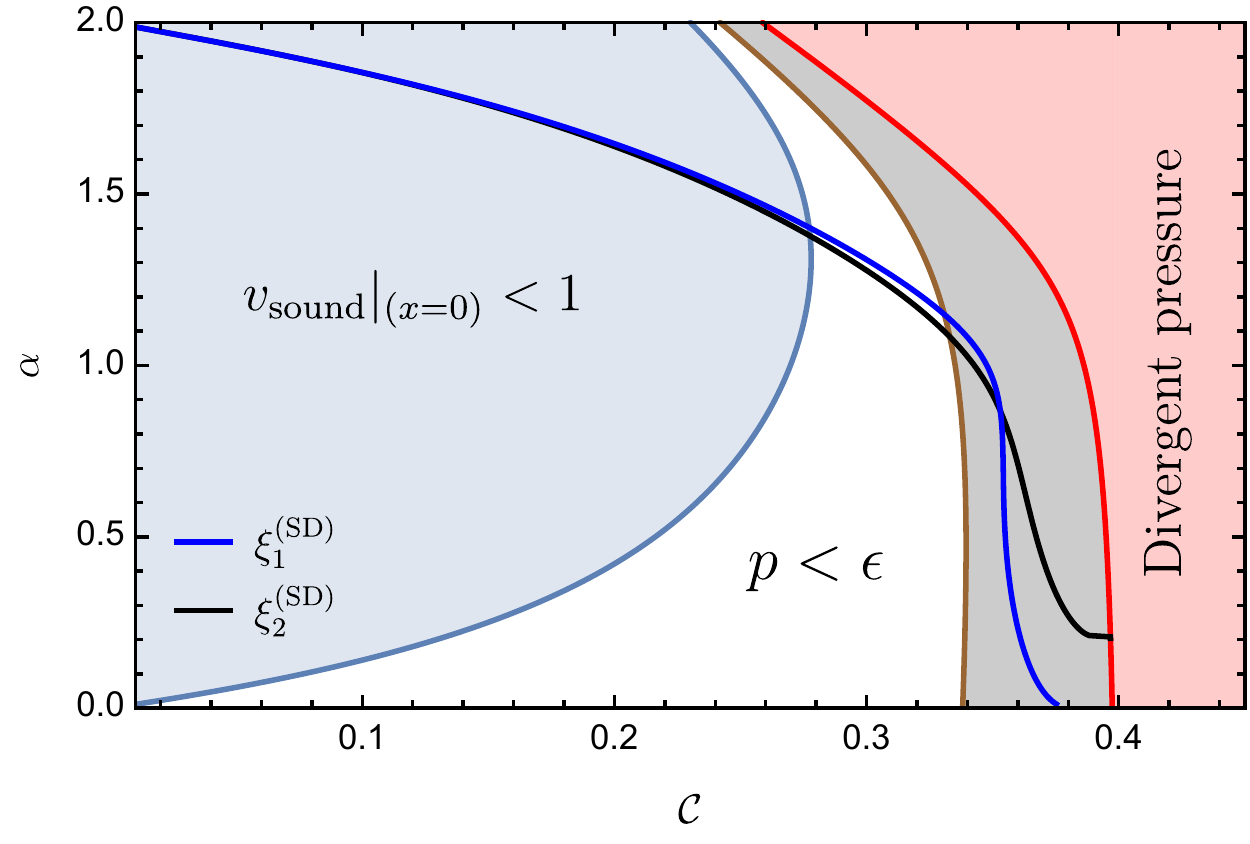}
\caption{Constraints on the compactness $\mathcal{C}$ and the parameter $\alpha$ for the modified T-VII solution. The solid light blue line indicates the causal limit as determined by the condition, $v_{\mathrm{s}}(x=0)\leq 1$, i.e., the central speed of sound must be less or equal to the speed of light (in units where $c=1$). The brown line corresponds to the limit set by the DEC. The solid red line illustrates the limit where the central pressure diverges. The blue and black solid lines indicate the limits on the stability domain (SD), given by the condition $\langle\gamma\rangle \geq \gamma_{\cri}$, as determined using the trial functions $\xi_1$ and $\xi_2$ (see Fig. \ref{fig:4} for details). Configurations with given parameters $(\mathcal{C}, \alpha)$ below these lines are dynamically stable.}
\label{fig:7}       % Give a unique label
\end{figure}

The stability regime for the modified T-VII solution is shown in Fig.~\ref{fig:4}, where we compare the effective adiabatic index $\langle\gamma\rangle$ [Eq.~\eqref{gammaeff}] and the critical value $\gamma_{\cri}$ [Eq.~\eqref{gammacr}]. The results depicted here correspond to the case $\alpha=0.7$ (left panel) and the original T-VII solution (right panel), and were determined using the trial function $\xi_2$. In ge\-ne\-ral we consider $\alpha$ in the whole range $\alpha\in[0,2]$, and we also used the trial function $\xi_{1}$; we will present all of our results in Fig.~\ref{fig:7}. The stability regime is given by the condition, $\langle\gamma\rangle \geq \gamma_{\cri}$ [Eq.~\eqref{stablec}]; thus the shaded region in Fig.~\ref{fig:4} corresponds to configurations which are dynamically unstable. The intersect point of the curves indicates the maximum compactness for the onset of instability. For instance, for the original T-VII solution we found that it becomes unstable for $\mathcal{C}\simeq 0.3427$, which corroborates the value reported by \cite{Moustakidis:2016ndw,Negi:1999grg}. Note that in the Newtonian limit $\mathcal{C}\to 0$, the critical adiabatic index approaches the well known value $4/3$.

\begin{figure*}[h]
\includegraphics[width=0.49\linewidth]{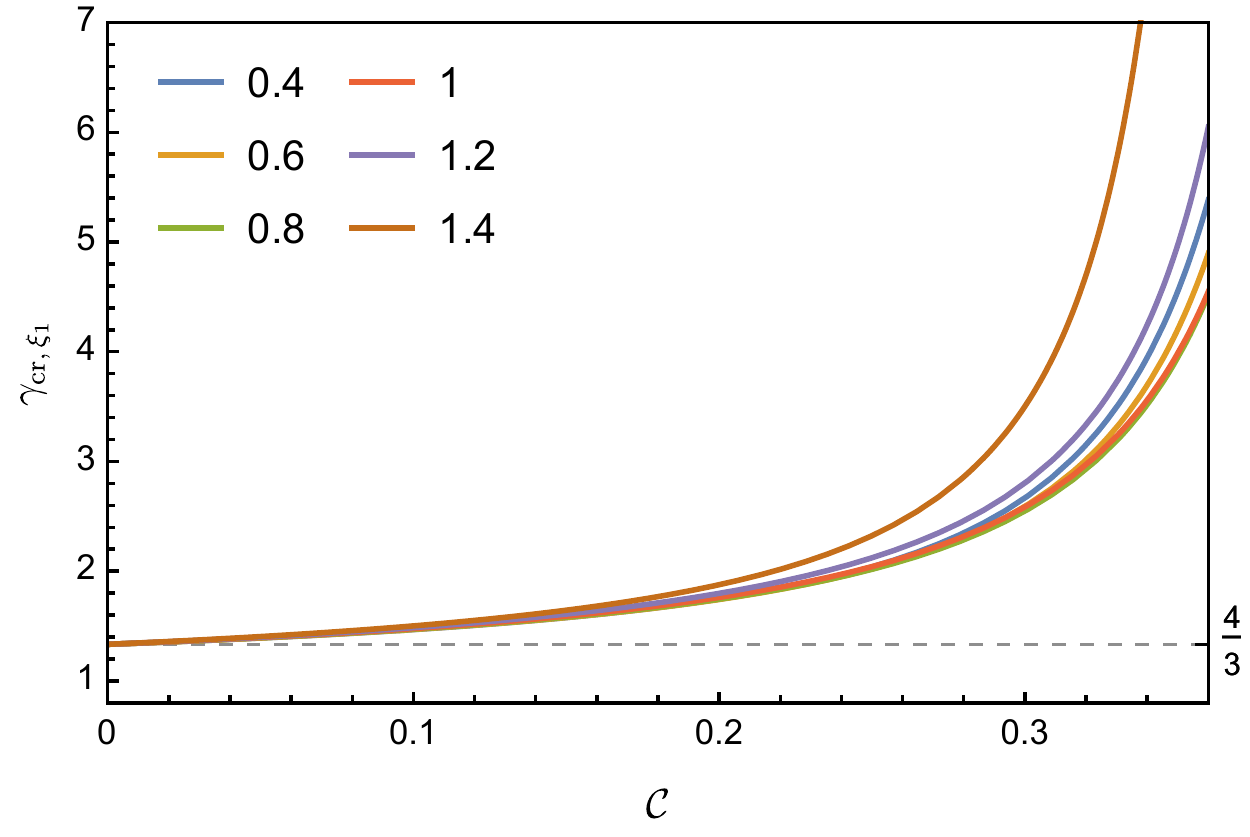}\hfill
\includegraphics[width=0.49\linewidth]{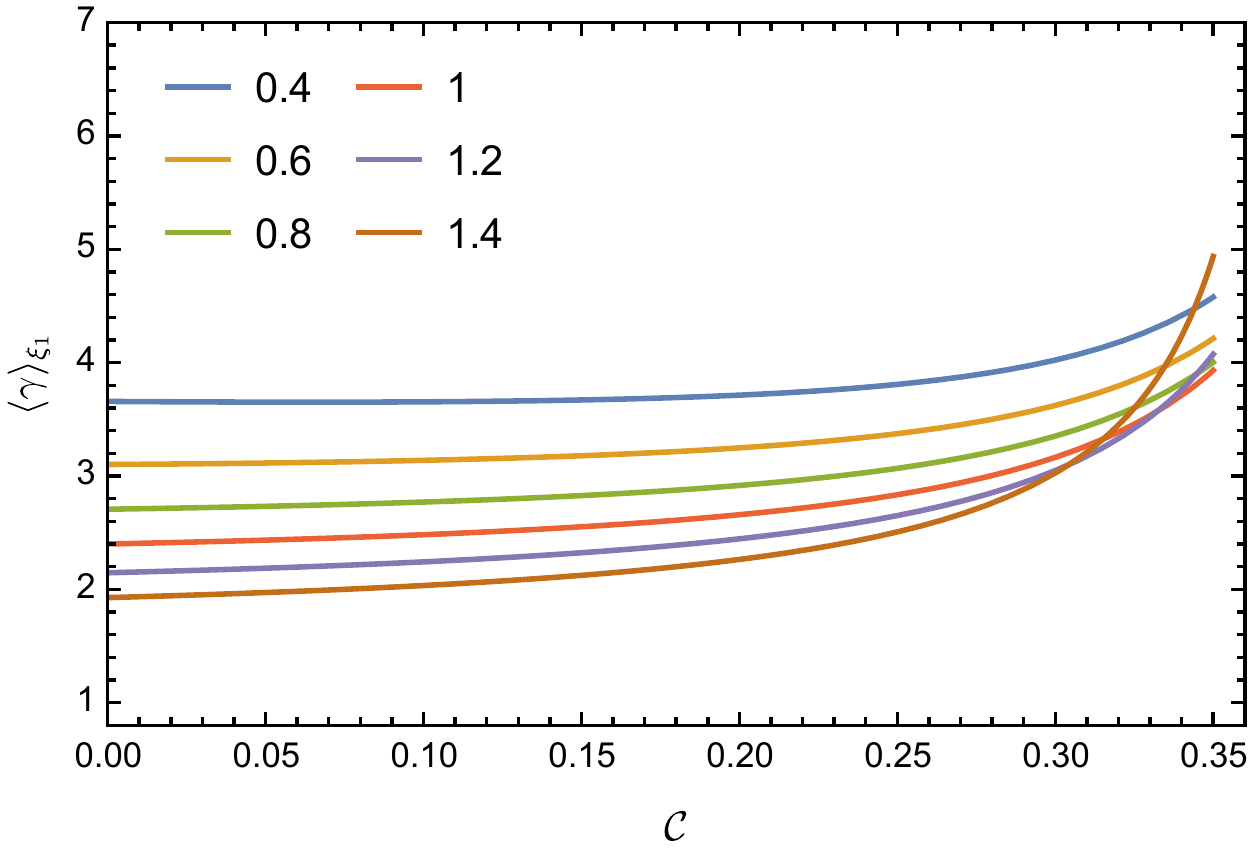}
\includegraphics[width=0.49\linewidth]{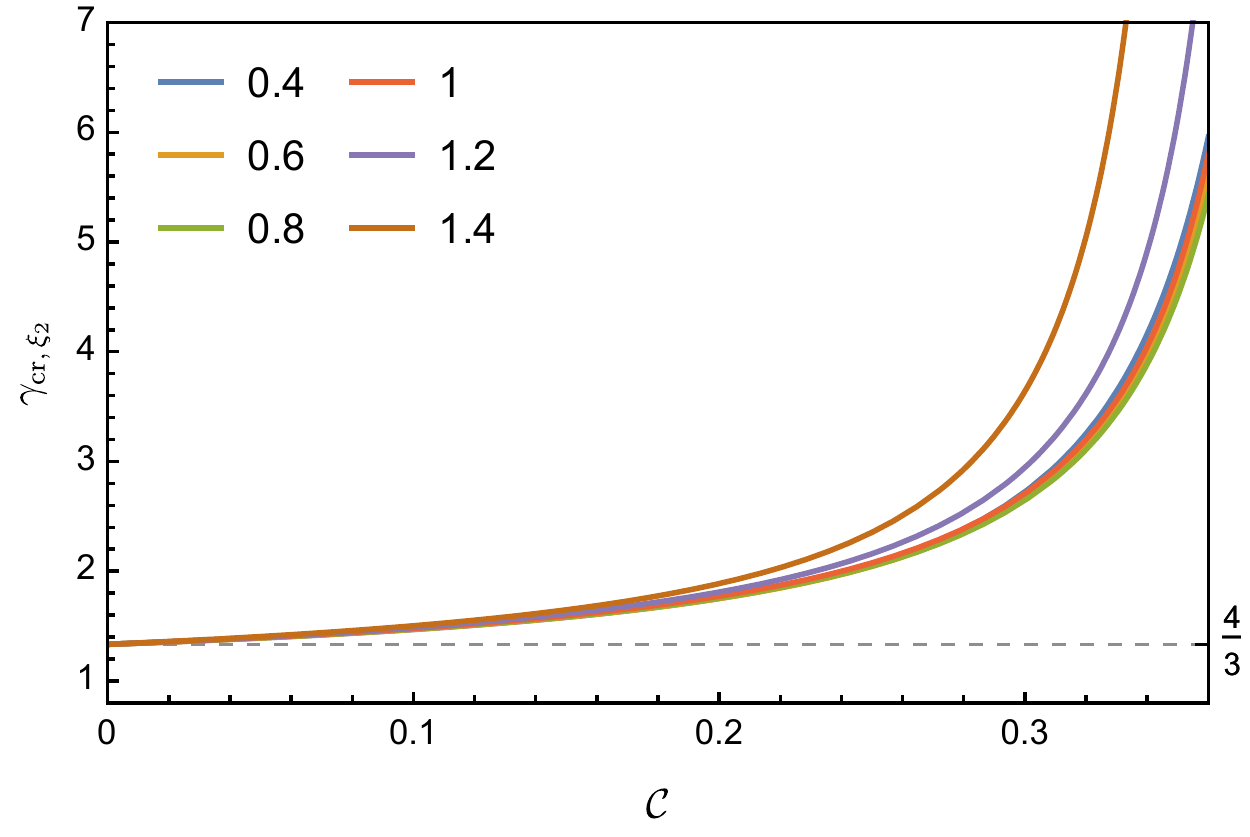}\hfill
\includegraphics[width=0.49\linewidth]{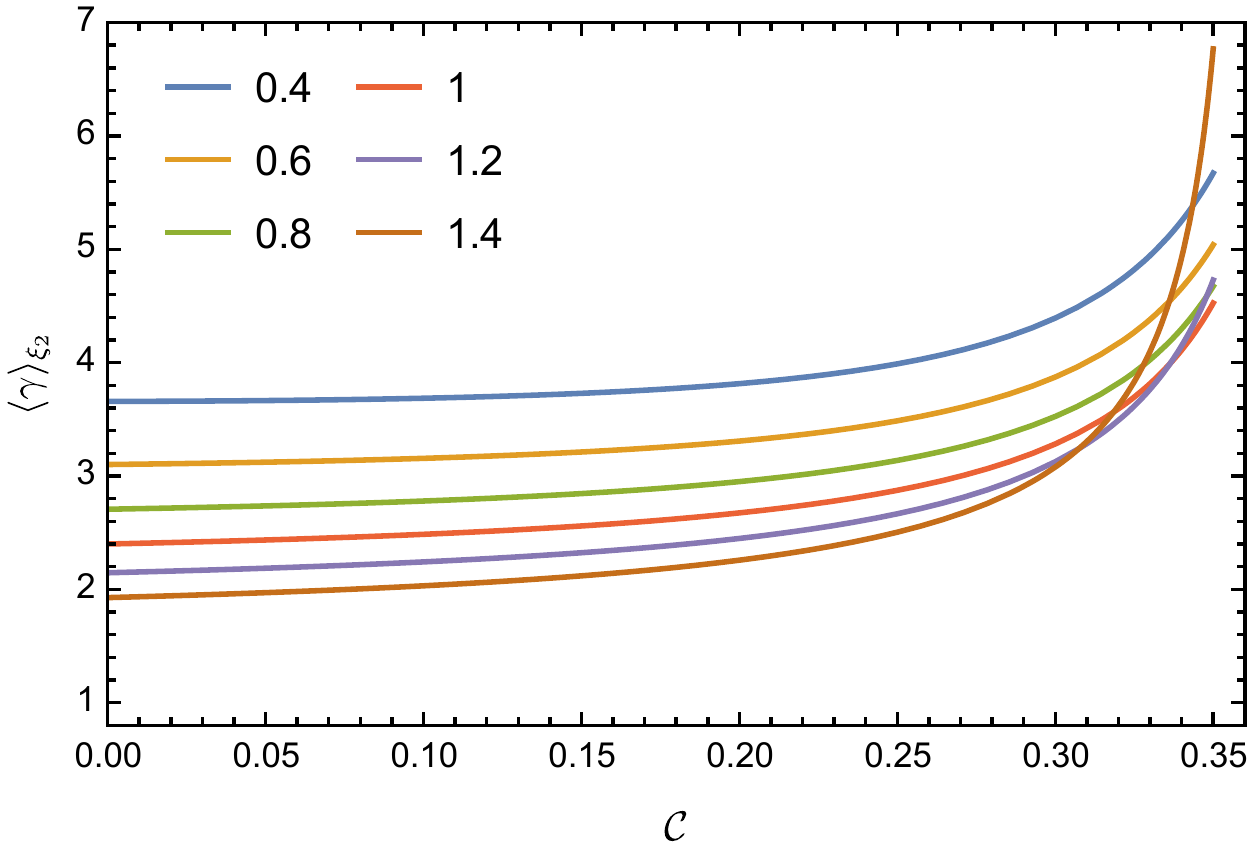}
\caption{Critical (top left panel) and effective (top right panel) adiabatic indices for the modified T-VII solution, as a function of the compactness, for some representative values of the parameter $\alpha$ as determined by using the trial function $\xi_{1}$. Bottom panels show the same quantities but determined using the trial function $\xi_{2}$.}
\label{fig:5}       % Give a unique label
\end{figure*}

In Fig.~\ref{fig:5} we display the critical adiabatic index $\gamma_{\cri}$, and the corresponding effective adiabatic index $\langle\gamma\rangle$, as a function of the compactness, for the modified T-VII solution. Here we consider several values of $\alpha\in[0.4,1.4]$. Note that in the Newtonian limit, all the cases approach to the expected value $\gamma_{\cri}=4/3$. We also observe that for configurations with the same compactness and $\alpha>1$, the critical adiabatic index rises as compared with the original T-VII solution. 

Some representative values of $\gamma_{\cri}$, for varying $\mathcal{C}$ and $\alpha$, are listed in Table \ref{tab:1}. For the original T-VII solution ($\alpha=1$), our results for the critical adiabatic index are in good agreement with those reported by \cite{Moustakidis:2016ndw}. It can be observed that for small values of the compactness $\mathcal{C}$, the $\gamma_{\cri}$ predicted by both trial functions is practically the same. For larger values of the compactness the trial function $\xi_2$ is ‘better' than $\xi_{1}$, in the sense that $\xi_2$ provides a lower value of $\gamma_{\cri}$. Note also that for fixed compactness, the increase in the parameter $\alpha$ raises the value of the critical adiabatic index. 

\renewcommand{\arraystretch}{1.3}
%\begingroup
\begin{table*}[ht]%\label{tab:1}
\begin{centering}
\begin{tabular}{c | cc | cc | cc | cc | cc | cc | cc }
\hline
\hline\noalign{\smallskip}
$\alpha$ & \multicolumn{2}{ c| }{0.4} & \multicolumn{2}{ c| }{0.6} & \multicolumn{2}{ c| }{0.8} & \multicolumn{2}{ c| }{1.0} & \multicolumn{2}{ c| }{1.2} & \multicolumn{2}{ c| }{1.4}\\
\hline%\hline
\noalign{\smallskip}\hline\noalign{\smallskip}
$\mathcal{C}$ & $\gamma_{\cri}^{\xi_1}$ & $\gamma_{\cri}^{\xi_2}$ & $\gamma_{\cri}^{\xi_1}$ & $\gamma_{\cri}^{\xi_2}$ & $\gamma_{\cri}^{\xi_1}$ & $\gamma_{\cri}^{\xi_2}$ & $\gamma_{\cri}^{\xi_1}$ & $\gamma_{\cri}^{\xi_2}$ & $\gamma_{\cri}^{\xi_1}$ & $\gamma_{\cri}^{\xi_2}$ & $\gamma_{\cri}^{\xi_1}$ & $\gamma_{\cri}^{\xi_2}$\\
\noalign{\smallskip}\hline\noalign{\smallskip}
0.007 & 1.3404 & 1.3404 & 1.3405 & 1.3405 & 1.3407 & 1.3407 & 1.3410 & 1.3410 & 1.3414 & 1.3414 & 1.3420 & 1.3420 \\
0.123 & 1.5124 & 1.5138 & 1.5135 & 1.5152 & 1.5169 & 1.5188 & 1.5238 & 1.5259 & 1.5364 & 1.5386 & 1.5594 & 1.5613 \\
0.169 & 1.6313 & 1.6350 & 1.6310 & 1.6355 & 1.6351 & 1.6404 & 1.6465 & 1.6524 & 1.6705 & 1.6766 & 1.7179 & 1.7232 \\
0.242 & 1.9785 & 1.9934 & 1.9637 & 1.9831 & 1.9620 & 1.9861 & 1.9849 & 2.0132 & 2.0552 & 2.0856 & 2.2259 & 2.2514 \\
0.279 & 2.3351 & 2.3679 & 2.2926 & 2.3368 & 2.2741 & 2.3308 & 2.3054 & 2.3743 & 2.4434 & 2.5201 & 2.8351 & 2.9022 \\
0.309 & 2.8581 & 2.9292 & 2.7587 & 2.8562 & 2.7005 & 2.8285 & 2.7393 & 2.9001 & 3.0118 & 3.2022 & 3.9396 & 4.1407 \\
%\noalign{\smallskip}
\hline
\hline
\end{tabular}
\end{centering}
\caption{The lower limits for $\gamma$ for dynamical stability of the modified T-VII solution, for various values of $
\mathcal{C}$ and $\alpha$. The $\gamma_{\cri}$ were evaluated with the trial functions $\xi_{1}$ and $\xi_{2}$.}
\label{tab:1}  
\end{table*}
%\endgroup

\begin{figure*}[ht]
\includegraphics[width=0.49\linewidth]{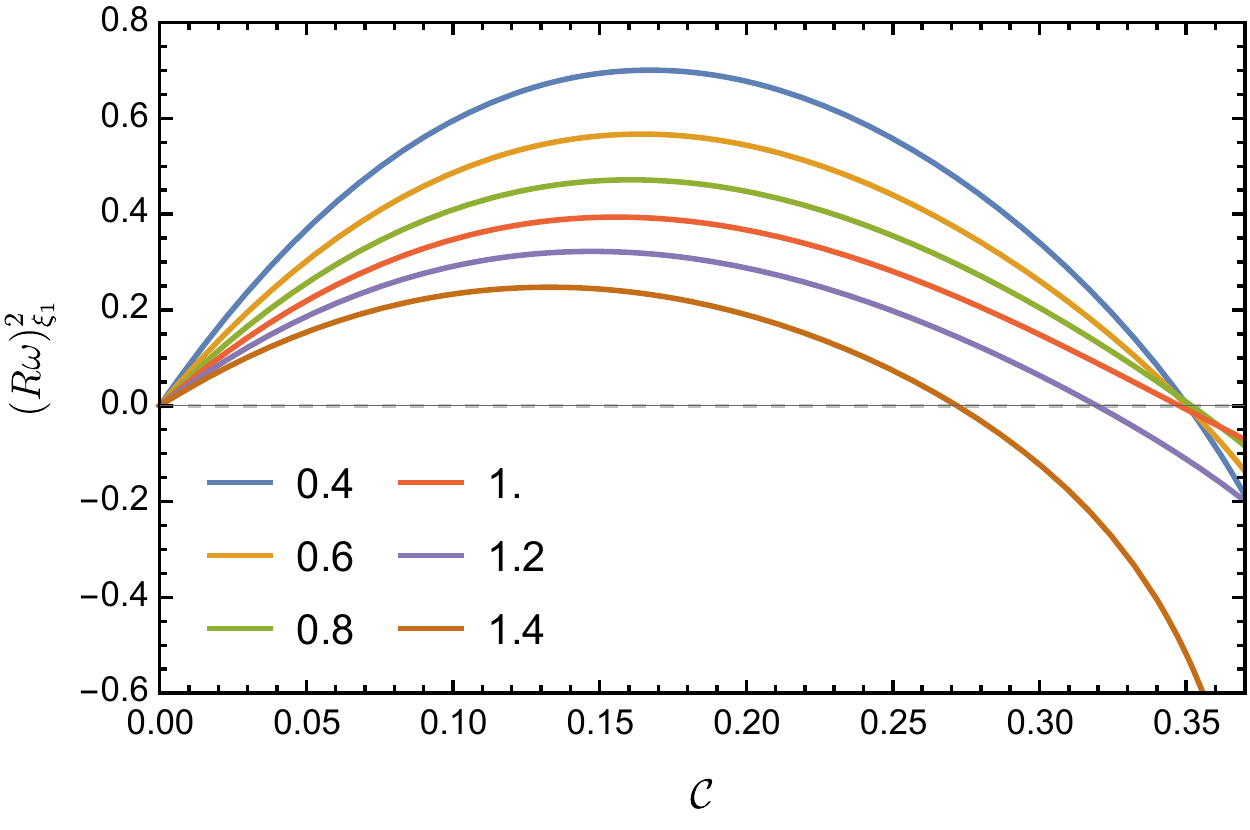}\hfill
\includegraphics[width=0.49\linewidth]{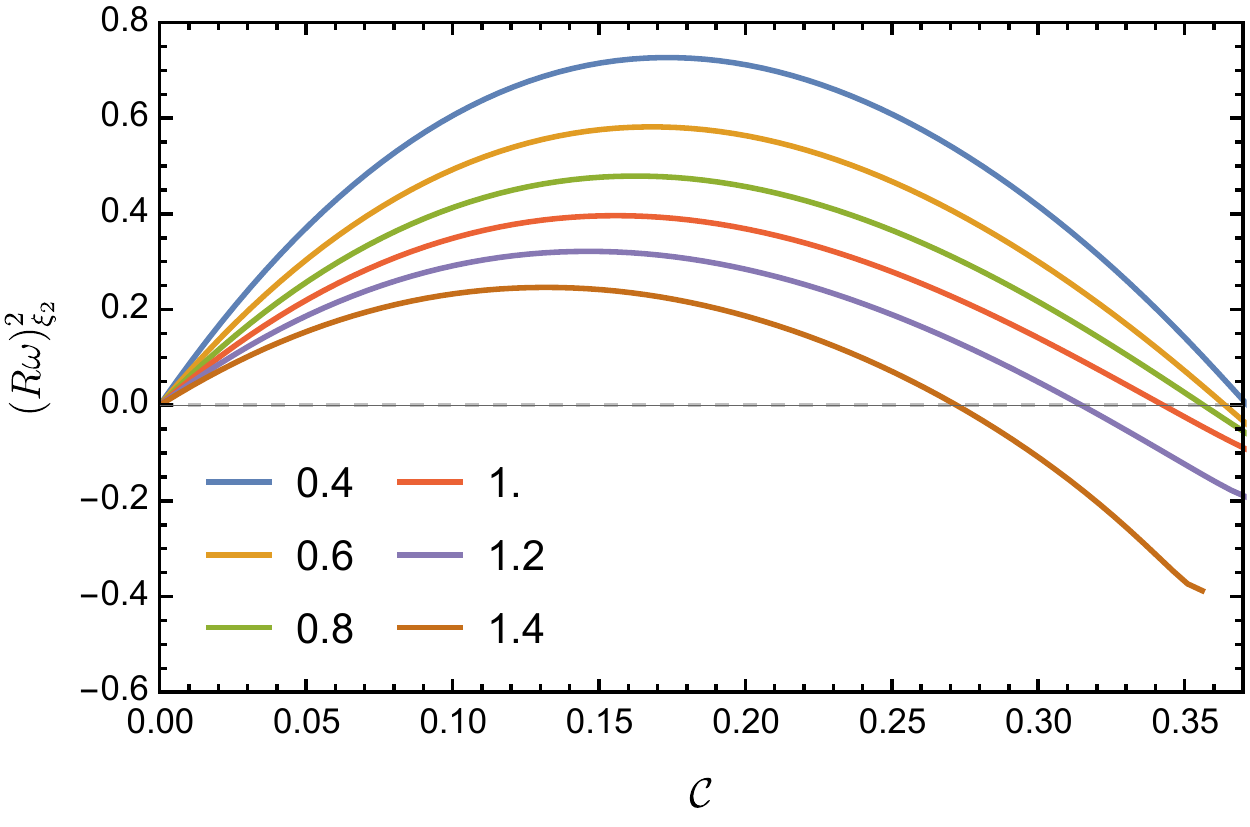}
\caption{The oscillation frequency in dependence of the compactness $\mathcal{C}$, for the modified T-VII solution, for various values of $\alpha$. The frequencies were computed from Eq.~\eqref{variat} by using the trial function $\xi_1$ (left panel) and the trial function $\xi_2$ (right panel).} 
\label{fig:6}       % Give a unique label
\end{figure*}

In Fig.~\ref{fig:6} we show the pulsation frequency $(R\omega)$ in dependence on the compactness, for the modified T-VII solution. We display the profiles for $\alpha\in[0.4,1.4]$. In the left (right) panel we show the frequencies computed using the trial function $\xi_1$ ($\xi_2$). For the original T-VII solution ($\alpha=1$) the maximum compactness for stability, as found using the trial function $\xi_2$, is $\mathcal{C}\simeq 0.3427$ which corroborates our result found previously using the stability condition $\langle \gamma \rangle > \gamma_{\cri}$. Both approaches follow from the variational principle [Eq. \eqref{variat}], so the agreement is expected.

\renewcommand{\arraystretch}{1.3}
%\begingroup
\begin{table*}[ht]%\label{tab:1}
\begin{centering}
\begin{tabular}{c | c | c | c | c}
\hline\hline%\noalign{\smallskip}
  & \multicolumn{4}{ c }{$\mathcal{C}$}\\%& \multicolumn{2}{ c| }{0.8} & \multicolumn{2}{ c| }{1.0} & \multicolumn{2}{ c| }{1.2} & \multicolumn{2}{ c| }{1.4} &  \multicolumn{2}{ c }{1.6}\\
%\hline%\hline
\noalign{\smallskip}\hline\noalign{\smallskip}
$\mathcal{\alpha}$ & $v_{s}(0)\leq 1$ & \text{Finite $p_c/\epsilon_\mathrm{c}$} & \text{Stability} & \text{DEC}\\
\noalign{\smallskip}\hline\noalign{\smallskip}
0.4 & 0.1950 & 0.3957 & 0.3651 & 0.3394  \\
0.6 & 0.2326 & 0.3940 & 0.3597 & 0.3392  \\
0.8 & 0.2556 & 0.3913 & 0.3549 & 0.3381  \\
1.0 & 0.2697 & 0.3861 & 0.3427 & 0.3351 \\
1.2 & 0.2769 & 0.3759 & 0.3131 & 0.3288  \\
1.4 & 0.2772 & 0.3570 & 0.2676 & 0.3171  \\
%\noalign{\smallskip}
\hline
\hline
\end{tabular}
\end{centering}
\caption{Limits on the compactness $\mathcal{C}$, for different values of the parameter $\alpha$, for the modified T-VII solution in order to satisfy the following conditions: (a) causality, (b) finite ratio $p_\mathrm{c}/\epsilon_\mathrm{c}$, (c) radial stability, (d) dominant energy condition.}
\label{tab:2}  
\end{table*}

The most important results of our analysis are summarized in Fig.~\ref{fig:7}, where we determine the stability and causality domains in the $(\mathcal{C},\alpha)$ parameter space, for the whole allowed domain in $\alpha\in[0,2]$. The solid blue line indicates the causality limit as determined by the condition that the speed of sound at the center of the configuration must be subluminal, $(\partial p/\partial\epsilon)_{x=0}<1$. The solid red line illustrates the limit where the central pressure diverges. The solid orange line indicates the limit determined by the DEC. Finally, the blue and black solid lines correspond to the stability domain (SD), as obtained via the trial functions $\xi_{1}$ and $\xi_{2}$, determined by the condition of dynamical stability discussed in Sec.~\ref{sec:3}.

A first observation is related to the limits established by the DEC. We note that for a given $\alpha$, the maximum compactness allowed by the DEC is less than the compactness set by the condition of finite central pressure. The whole region where the DEC is violated is shaded in Fig.~\ref{fig:7}. Even though in this region the central pressure is finite, our results indicate that there would not be possible to find a configuration in this region which could represent a model of a NS in this framework.

Regarding the stability, we observe that the trial function $\xi_2$ puts stricter constraints on the maximum compactness for stability, for a given $\alpha$, as compared to the trial function $\xi_1$, at least up to $C\sim 0.35$ (see Fig. \ref{fig:7}). For low values of $\mathcal{C}$ there is good agreement in the stability limits predicted by both trial functions; however, as the compactness grow, the differences also grow. 

%************************************************************************************************
%************************************************************************************************

\section{Conclusions and discussions}
\label{sec:5}

In this paper we studied the stability, under radial oscillations, of the recently proposed modified Tolman VII solution. For this purpose we employed the well-established linearized analysis of time-dependent radial perturbations developed by Chandrasekhar \cite{Chandrasekhar:1964zz}. We solved the Chandrasekhar ‘pulsation' equation and determined the critical adiabatic index $\gamma_{\cri}$, for the onset of instability, for different values of the parameters $\alpha$ and compactness $\mathcal{C}$. First of all, we confirmed the results reported by \cite{Negi:1999grg, Negi:2001ApS, Raghoonundun:2015wga, Moustakidis:2016ndw} on the dynamical stability of the original Tolman VII solution ($\alpha=1$) for a wide range of compactness. We extended our analysis for values of $\alpha$ in the whole range allowed by the solution, i.e., $\alpha\in[0,2]$. Our study shows that the modified T-VII solution is stable against radial oscillations for a considerable range of values of the parameters ($\mathcal{C},\alpha$), thus supporting the physical viability of this solution as an approximate realistic model for the interior of neutron stars. 

Let us comment on the methods we used here. A different approach to study the dynamical stability of relativistic spheres, rely on energy considerations. The connection between the extremum mass and the normal modes is discussed in the classical textbooks \cite{Harrison:1965,Shapiro:1983du}. Since the energy density profile of the modified T-VII solution is fixed by the initial assumption, the adaptation of this method in this situation is not suitable.

%************************************************************************************************
%************************************************************************************************

\begin{acknowledgements}
We are thankful to the anonymous referee for the valuable comments and suggestions to improve this manuscript. We also acknowledge the support of the Institute of Physics and its Research Centre for Theoretical Physics and Astrophysics at the Silesian University in Opava.
\end{acknowledgements}

%************************************************************************************************
%************************************************************************************************

%\bibliography{main_rv}
%

\end{document}